\documentclass[pdftex,5p,times]{elsarticle}
\usepackage{hyperref}
\usepackage{url}
\usepackage{amsmath}
\usepackage{mathtools}
\usepackage[numbers]{natbib}
\usepackage{caption}
\usepackage{subcaption}
\usepackage{alltt}
\usepackage{color}
\usepackage[retainorgcmds]{IEEEtrantools}
\usepackage{textcomp} 
\usepackage{threeparttable} 
\usepackage{rotating}

\usepackage{stfloats}

\usepackage{amssymb}

\biboptions{sort&compress}

\journal{Journal of Manufacturing Processes}

    \renewcommand{\dbltopfraction}{0.99}	
    \renewcommand{\dblfloatpagefraction}{0.7}	

\usepackage{enumitem}

\usepackage[]{geometry}

\begin{document}

\begin{frontmatter}



\title{Two-dimensional thermal finite element model of directed energy deposition:\\
  matching melt pool temperature profile with pyrometer measurement}


\author[lab0]{Bohumir Jelinek}
\author[lab0,lab1]{Joseph Young}
\author[lab0,lab1]{Matthew Dantin}
\author[lab0,lab1]{William Furr}
\author[lab0]{Haley Doude}
\author[lab0,lab1]{Matthew W.\ Priddy}
\address[lab0]{Center for Advanced Vehicular Systems,
    Starkville, MS 39759, USA}
\address[lab1]{Mechanical Engineering Dept.,
    Mississippi State University, MS 39762, USA}

\begin{abstract}
    An open source two-dimensional (2D) thermal finite element (FE) model of the Directed Energy Deposition (DED) process is developed using the Python-based FEniCS framework. The model incrementally deposits material ahead of the laser focus point according to the geometry of the part. The laser heat energy is supplied by a Gaussian-distributed heat source while the phase change is represented by increased heat capacity around the solidus-liquidus temperature range.
    Experimental validation of the numerical model is performed by matching with the melt pool temperature measurements taken by a dual wavelength pyrometer during the build process of a box-shaped Ti-6Al-4V part with large geometrical voids.
    Effects of large geometrical voids on the melt pool shape and maximum melt pool temperature are examined. Both the numerical and experimental data show an increase in the melt pool size and temperature during deposition above large voids. The trailing edge of the melt pool's temperature profile obtained using the developed numerical model closely matches pyrometer measurements.
\end{abstract}


\end{frontmatter}


\section{Introduction}

Additive manufacturing (AM) is a flexible technology allowing incremental production of components with complex geometries from a wide range of materials. The growth in application of this technology  accelerated in recent years, attracting attention of both industry and academic research \cite{Kruth1998,astm2010:am,Guo2013,Mellor2014,Frazier2014,Rubenchik2018,Ngo2018,Loh2018}.
Compared to traditional formative and subtractive manufacturing methods, AM exhibits shorter production times. AM can directly follow Computer Aided Design (CAD) model of the part and build a near net shape product. Applications of AM technology include cladding, tool repair, and a production of functionally graded materials.

Feedstock material for AM of metallic parts assumes the form of a wire or a powder. Melting and bonding of the feedstock is performed by a moving heat source, which can be an electron beam or a laser. Two common powder-based AM techniques are Powder Bed Fusion (PBF) and Directed Energy Deposition (DED). In PBF, a full layer of powder is spread on a build platform, followed by melting/sintering of a chosen portion of the surface layer to form the part. The platform is then lowered, a full layer of powder is spread and selectively melted, forming a new layer of the part being built. In laser based DED, the powder is blown through nozzles into the laser beam focus point on the build surface. The nozzles and the laser source are mounted on the deposition head which moves according to a preprogrammed deposition path. Both PBF and DED AM processes are characterized by phenomenally rich thermal history because localized cyclical heating produces high thermal gradients, fast heating/cooling rates, and multiple melting/solidification cycles.

Complex thermal history of additively manufactured parts, especially those made of metallic alloys with microstructure depending on melting and solidification rates, results in strong and difficult-to-predict dependencies of the product quality on manufacturing process parameters \cite{Gu2012,Thompson2015,Bian2015,Seifi2016,Thompson2016}. Uncertainties in quality prevent wider AM deployment. Thermal history depends on the parameters of the build process, which are laser power, deposition path, part geometry and dimensions, initial and build environment conditions. Suboptimal or non-uniform thermal histories produce more defects, such as heterogeneous microstructure \cite{Wang2006,Bian2015,Hernandez-nava2016:defects}, porosity \cite{kobryn2001mechanical,Galarraga2016,wang2009experimental,Liu2015}, and residual stresses \cite{Rangaswamy2005,Zheng2008,Thompson2015,liu2015residual}, leading to part distortion and degradation of mechanical properties. Components produced via DED commonly require post-manufacturing processes (e.g., machining or hot isostatic pressing) to account for the presence of process induced defects.

The ability to predict and control thermal history in AM would have a significant impact on improving the as-built part and reducing cost.
Two- and three-dimensional finite element (FE) based numerical modeling of the directed energy deposition of Ti-6Al-4V parts have been subject to multiple research efforts \cite{Yin2008,Denlinger2014,denlinger2015:res,Heigel2015,Teng2017}. \citet{Denlinger2014} and \citet{Heigel2015} developed and validated a thermo-mechanical model of electron beam and laser deposition of a thin Ti-6Al-4V wall using FE solver as implemented in CUBIC (PAN Computing, LLC \cite{cubic2}).
Similarly to the work presented here, several studies examined the details of the melt pool profile during the Ti-6Al-4V deposition. The temperature profile and melt pool depth in laser powder bed fusion of Ti-6Al-4V was analyzed using a finite difference method as implemented in Matlab{\textregistered} \cite{matlab2} by \citet{Criales2017}.  \citet{Cheng2014} used Abaqus \cite{abaqus2} software with custom DFLUX subroutine implementing a volumetric heat source to examine the temperature profile and melt pool size during Ti-6Al-4V electron beam additive manufacturing (EBAM)
process. 
\citet{peyre2008} and \citet{laBatut2017} compared experimental, analytical, and numerical predictions of a temperature field during direct metal deposition of Ti-6Al-4V with their numerical models implemented in COMSOL Multiphysics{\textregistered} \cite{comsol2}.
\citet{Vincent2018} implemented thermal-fluid numerical simulation
using the simpleFOAM solver of the OPenFOAM suite \cite{Weller1998} to
predict melt pool width and height for a single straight track
deposition on a flat surface. The prediction was achieved by
iterative optimization of analytically defined track geometry with the
width and height of the melt pool assumed equal to the width and height
of the track while the track cross-section shape was approximated by a
circular segment. Simulations show a good agreement with 
experiment with the largest relative error of 21.4\% in the track
height for a steady state of the deposition on a homogeneous flat
substrate.
\citet{Lu2019} used an in-house research software COupled MEchanical and Thermal (COMET) \cite{cervera2002comet,Michele2017} to characterize residual stress and distortion in rectangular and S-shaped Ti-6Al-4V parts built using DED. The largest average temperature error at all thermocouple
locations was lower than 13\%.
Recently, a fast semi-analytical thermal model of DED process
including phase change and gas flow effects was developed by
\citet{Weisz-Patrault2020} using Scilab~\cite{scilab}. The model,
applicable to single-track builds with geometries that do not vary in
vertical direction, was compared with pyrometer measurement at two
part locations achieving a good agreement with only 2.6\% average
error.
As seen, the simulations of the DED process were performed using commercial software, while some used in-house research codes.


\begin{figure}[!bp]
  \center
  \includegraphics[trim=0in 0in 0in 0in, clip,width=1\columnwidth]{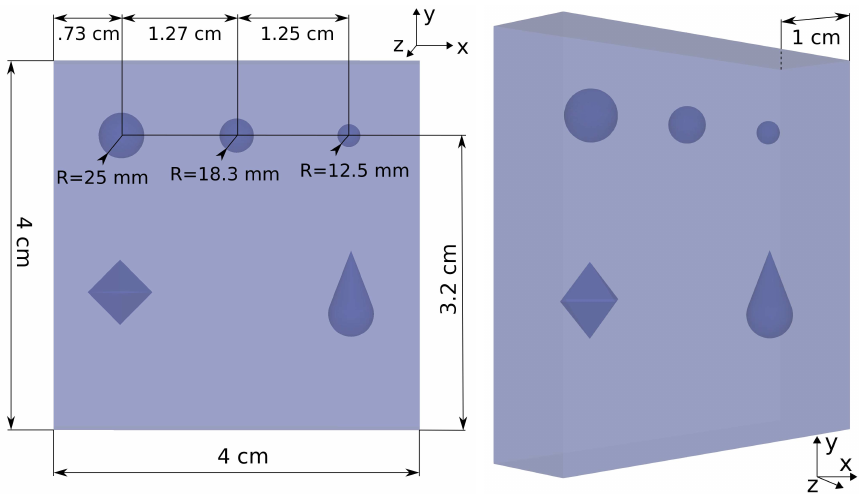}
  \caption{Front and diagonal view of the part featuring large voids.}
  \label{fig:geom}
\end{figure}

This work introduces an open source two-dimensional (2D) FE numerical model of heat transfer during the DED process developed within an open-source FEniCS \cite{AlnaesBlechta2015a} framework.
Python source code implementing the numerical model and resulting animations are available at \href{https://gitlab.com/bohumir1/2d-heat-ded}{GitLab} under LGPL-3.0 license.
 Unlike the bulk of published work addressing PBF, this work investigates DED process, specifically Laser Engineered Net Shaping (LENS{\textregistered}, Optomec). The PBF additive manufacturing process spreads a full layer of new powder in one step. Selected portions of the full layer are melted only after the complete layer is spread. The PBF melt pool is not constantly exposed to cold, freshly deposited material. Contrary to PBF, the DED process \textit{continuously} deposits cold material directly into the melt pool. The model of the LENS DED process therefore needs to implement incremental material deposition. Modeling of incremental DED process comes with intense numerical challenges. First is the convergence issue due to thermal gradients arising from the cold material being continuously deposited into the melt. Aside from convergence issues, incremental deposition requires continuous adding of new elements, modifying boundary conditions, and rebuilding the system of equations repeatedly during the deposition of a single line. While the finite element model of PBF rebuilds the system of equations only once per each deposited layer, the incremental deposition implemented in this work performed 1580 rebuilds of the system of equations during the deposition of a single layer.
High computational demands and the divergence of Newton solver's iterations caused by the cold material being deposited directly into the melt pool are the main challenges addressed in the newly developed FEA DED model, affording a detailed look at the melt pool temperature profile.
%

%
%
The earlier works do not examine details of the melt pool temperature profile. Further, they do not look at the variation of the melt pool temperature profile above large geometrical voids.
The newly developed thermal model of incremental DED process is applied to examine the temperature distribution during direct laser deposition of a Ti-6Al-4V alloy part with the presence of large geometrical voids.  For this study, a 2D model was used in lieu of a 3D model because of the increased computational efficiency, ease of mesh refinement around the voids, and the ability to readily change the void's size and location. A rectangular prism shaped part with a geometry incorporating large voids was designed and then built using the Laser Engineered Net Shaping (LENS{\textregistered}), which is a popular flavor of DED developed at Sandia National Laboratories in the 1990s \cite{atwood1998laser}. The melt pool temperature during the build process was monitored by a dual wavelength pyrometer as described in \cite{Marshall2016}. The finite element model was then constructed and correlated with temperature distribution from the build experiment.

\section{Part geometry and the deposition process}

The geometry of the part examined in this study is shown in Figure \ref{fig:geom}.
The dimensions of the complete block are 4~cm (width, along $x$-direction) $\times$ 4~cm (height, along $y$-direction) $\times$ 1~cm (depth, along $z$-direction).
Spherical voids were located along the lateral mid-section of the part, with centers at the distances 0.73~cm, 2.0~cm, and 3.25~cm from the left edge, and 3.2~cm from the bottom edge.  Additionally, diamond and tear-drop shaped voids were part of the build experiment but not included in simulations.
The radii of the spherical voids were 0.25~cm, 0.183~cm, and 0.125~cm.

The part was built by depositing 79 layers of Ti-6Al-4V powder. The direction of deposition was rotated by 90 degrees for each successive layer. Odd layers were formed by 19 tracks deposited along the $x$-direction. Even layers were formed by tracks deposited in the $z$-direction. The 2D numerical model was compared with the deposition of the 10th track of the 69th layer, chosen as a center line of 19 tracks of the layer 69.

\begin{figure}[!tbp]
  \center
  \includegraphics[trim=0in 0in 0in 0in, clip,width=1\columnwidth]{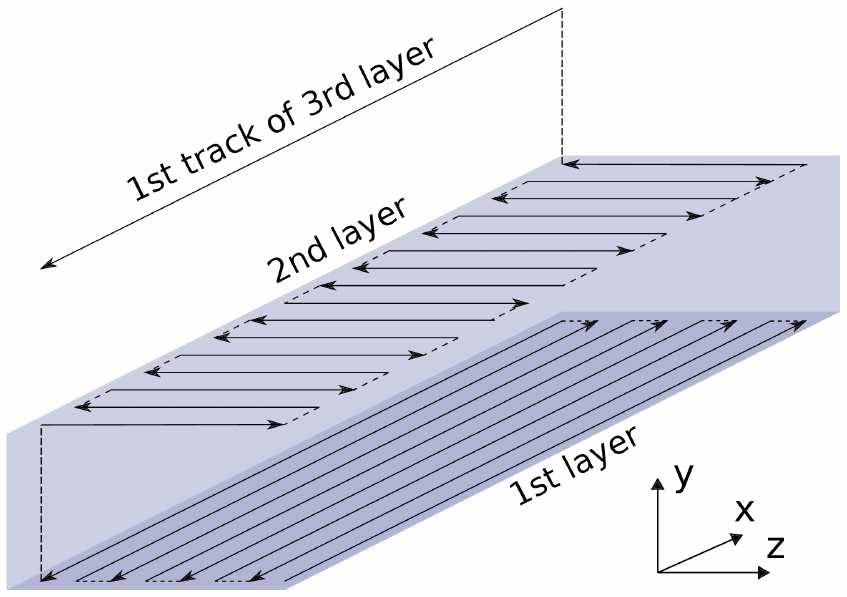}
  \caption{Deposition direction of successive layers.}
  \label{fig:depo}
\end{figure}

\section{Numerical representation}

\subsection{FEniCS framework}

Open source finite element suite FEniCS \cite{LoggMardalEtAl2012a} was used to build a numerical heat transfer model of the DED process. The FEniCS framework \cite{AlnaesBlechta2015a} consists of a collection of interoperable software components, including Dynamic Object-oriented Library for FINite element computation DOLFIN \cite{LoggWells2010a,LoggWellsEtAl2012a}, FEniCS Form Compiler FFC \cite{LoggOlgaardEtAl2012a}, FInite element Automatic Tabulator FIAT \cite{Kirby2004a,Kirby2012a}, inlining module \textit{Instant} built on top of Simplified Wrapper and Interface Generator SWIG \cite{Beazley:1996:SEU:1267498.1267513} and Python package \textit{distutils} \cite{distutils}, Unified Form-assembly Code UFC \cite{AlnaesLoggEtAl2009a,AlnaesLoggEtAl2012a}, Unified Form Language UFL \cite{Alnaes2012a}, and a mesh generation component \textit{mshr} utilizing tetrahedral mesh generator TETGEN~\cite{si2015tetgen} and Computational Geometry Algorithms Library
CGAL \cite{cgal:eb-18b} as mesh generation backends.

\subsection{Heat transfer model}

Heat transfer is governed by the heat equation
\begin{equation}
  \rho c_p(T) \frac{\partial T}{\partial t}  - k \Delta T = 0,
\end{equation}
where $\rho$ is the density, $T$ is the temperature, $t$ is the time, and $k$ is the thermal conductivity. The specific heat $c_p$ was increased for the temperature range between solidus $T_S$ and liquidus $T_L$ to account for the latent heat of melting $L$ as follows

\begin{figure}[!b]
  \center
  \includegraphics[trim=0in 0in 0in 0in, clip,width=1\columnwidth]{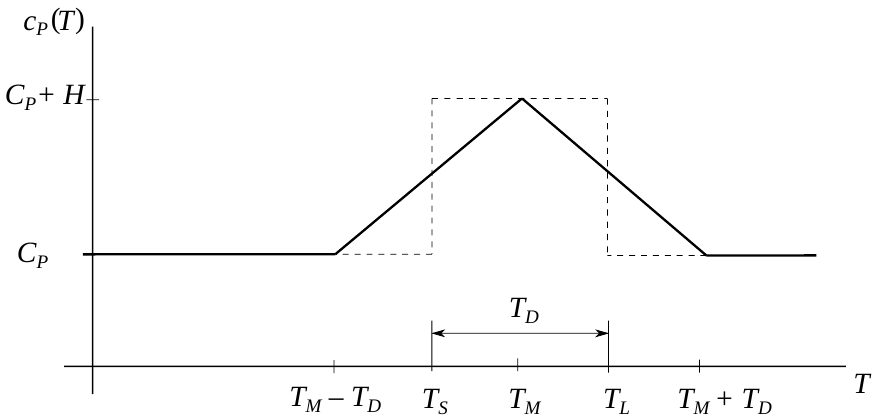}
  \caption{Representation of the latent heat.}
  \label{fig:lh}
\end{figure}

\begin{equation}
c_P(T) = 
	\begin{dcases}
	C_P + H\left(1-\frac{|T-T_M|}{T_D}\right) & \text{if  }  |T-T_M| < T_D,\\
	C_P & \text{otherwise}.
    \end{dcases}
    \label{eq:sh}
\end{equation}
The temperatures $T_M$, $T_D$, and the increment $H$ of specific heat are defined as
\begin{equation}
\begin{split}
  T_M & = (T_L + T_S)/2\\
  T_D & = T_L - T_S\\
    H & = \frac{L}{T_D}.
\end{split}
\end{equation}
The temperature dependence of $c_p$ (Equation \ref{eq:sh}) is illustrated in Figure \ref{fig:lh}.
The triangle-shaped continuous function improved the stability of the FE solver in comparison with a step function alternative deployed by \citet{Piekarska2010ApplicationOA}.


\begin{table*}[!b]
\centering
\begin{threeparttable}
\small
  \begin{tabular}{|l|c|c|c||l|c|c|c|}
    \hline
Material Parameter                   & Symb.        & Value  & Units &               Process Parameter                    & Symb.        & Value  & Units \\        \hline \hline 
Thermal conductivity                 & $k$          & 7.2    & W/m/K &               Laser power                          & $P$          & 300     & W\\                    
Specific heat of solid               & $C_p$        & 560    & J/kg/K &              Laser power abs. coef.              & $\alpha$     & $9.6\times10^{-3},7.2\times10^{-3}$\tnote{*} & - \\
Latent heat                          & $L$          & $3.65 \times 10^5$ & J/kg &    Laser beam width                     & $w_{b}$      & $0.5\times10^{-3}$  & m\\       
Density of material                  & $\rho$       & 4420   & kg/m$^3$ &            Layer thickness                      & $h$          & $0.508\times10^{-3}$  & m\\     
Thermal emissivity                   & $\epsilon$   & 0.54   & - &                   Deposition velocity                  & $v_{b}$      & $16.93\times10^{-3}$  & m/s\\   
Convective heat transfer coef.       & $h$          & 30     & W/m$^2$/K &           Initial/bottom temperature                  & $T_{i}$      & 127, 30, 427\tnote{*} & \textdegree C\\        
Solidus-liquidus range         & $T_{s}-T_{l}$ & 1600$-$1650 & \textdegree C &          External temperature                 & $T_{e}$      & 30     & \textdegree C\\  \hline      
   \end{tabular}
  \begin{tablenotes}
  \item[*] As described in Section~\ref{sec:simulations}, 
    $\alpha = 9.6\times10^{-3}$ in cases 1 and 2, and $7.2\times10^{-3}$ in case 3, while
    $T_i = 127$~\textdegree C in case 1, $30$~\textdegree C in case 2, and $427$~\textdegree C in case 3
  \end{tablenotes}
\end{threeparttable}
\caption{Ti-6Al-4V material properties and parameters of the deposition process.\label{tab:params}}
\end{table*}

\subsection{Boundary conditions}

\begin{figure*}[!t]
  \center
  \includegraphics[trim=0in 0in 0in 0in, clip, width=1.0\textwidth]{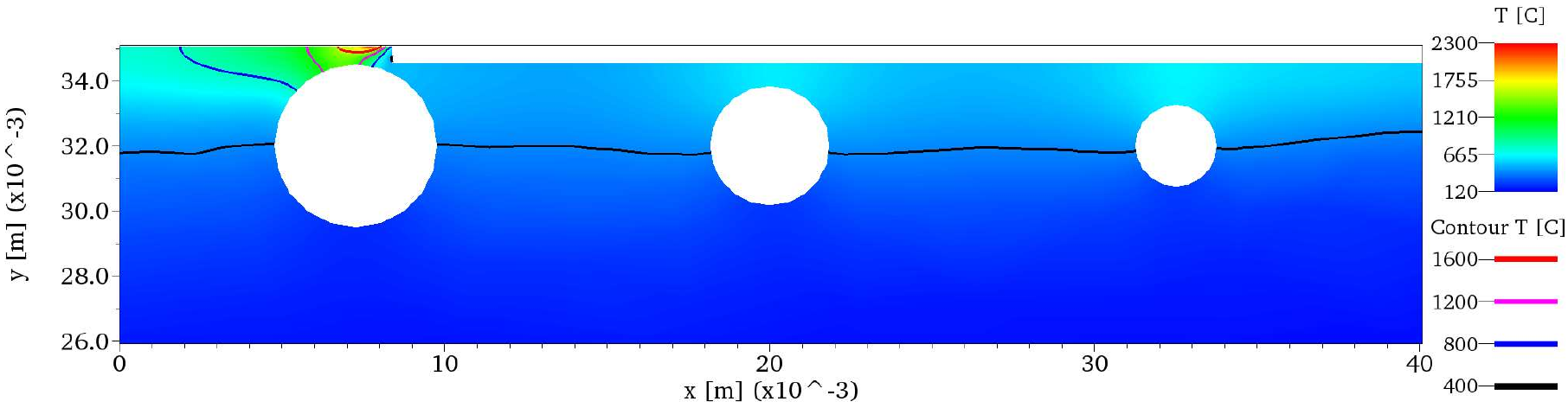}
  \caption{Simulated temperature within upper portion of the part when depositing above the largest void. Mesh in the top-most portion of the simulation is displayed in Figure~\ref{fig:over}. Note that the simulation domain extends to $y=0$.}
  \label{fig:upper}
\end{figure*}

On the bottom boundary ($y=0$ cm), the temperature is kept constant
\begin{equation}
  T = T_i.
\end{equation}
On the left ($x=0$ cm) and right ($x=4$ cm) boundaries, and on the surface of voids, a convection boundary condition is imposed
\begin{equation}
  -k\frac{\partial T}{\partial x} = h(T - T_e ),
\end{equation}
where $h$ is the convective heat transfer coefficient.
On the top boundary, the heat supplied by the laser, $p(x)$, as well as heat release by convection and radiation are assumed
\begin{equation}
  -k\frac{\partial T}{\partial x} = p(x) + h(T - T_e ) + \epsilon \sigma (T^4 - T_e^4),
\end{equation}
where $\sigma$ is the Stefan-Boltzmann constant $(5.67\times10^{-8}$ Wm$^{-2}$K$^{-4})$ and $\epsilon$ is the thermal emissivity of Ti-6Al-4V. The heat supplied by the laser is represented by a Gaussian heat source \cite{Yilbas1997}
\begin{equation}
  p(x) = A \, \text{exp} \left( -\frac{1}{2} \frac{ (x-x_0)^2 }{ (w_b/2)^2 } \right),
    \label{eq:gauss}
\end{equation}
where
\begin{equation}
  A = \alpha \frac{P}{2\pi(w_b/2)^2}
  \label{eq:gmag}
\end{equation}
is the magnitude of the power distribution function, $\alpha$ is the effective heat absorption coefficient for laser power $P$, $x_0$ is the displacement of the laser focus point, and $w_b$ is the laser beam width. Because $A$ is the magnitude of the 1D Gaussian function (Equation \ref{eq:gauss}) and the total laser power $P$ is distributed over the 2D surface, Equations \ref{eq:gauss} and \ref{eq:gmag} are chosen to represent a section of 2D Gaussian distribution with the coordinate $y=0$. The DED process parameters and calibrated effective laser absorption coefficient $\alpha$ are presented in Table~\ref{tab:params}. Other 2D models \cite{Wang2006,Yin2008} calibrate the 1D magnitude $A$ instead of $\alpha$ without listing its value. The calibration of $\alpha$ in the present study provides more physical information. The Ti-6Al-4V material properties and the deposition process parameters are listed in Table \ref{tab:params}.


\subsection{Model of the deposition process}

The new material at temperature $T_i$ is deposited at $\Delta x = 2 \, w_b$  distance ahead of the laser beam focus point to assure the stability of Newton's FE solver. A single column of material with temperature $T_i$ is deposited at a time. The time intervals between depositions of successive columns correspond to the laser velocity $v_b$ as listed in Table~\ref{tab:params}.

\subsection{Finite element mesh}
The triangular FE mesh with linear elements used in this work was constructed using the \textit{mshr} module of FEniCS. 
The mesh was refined within three layers of the simulated deposition process (Section~\ref{sec:simulations}) as well as on the edge of the voids.
The largest and the most frequent mesh element within the deposited layers was a right triangle with two sides of 25.4~$\mu$m.
This refinement accommodated the laser beam width of 500~$\mu$m and also allowed material deposition increments of 25.4~$\mu$m. 
Each of the deposited layers was split into a grid of 1580 columns and 20 rows for deposition purposes.  
A complete column at the leading edge of the layer was deposited in each deposition step, resulting in a deposition layer with a thickness of 508~$\mu$m.
Twice-coarser mesh refinement was applied within the layer under the three deposited layers to accommodate the heating pass.
The total number of elements in each simulation was 206,012, with a total of 104,212 vertices.
The mesh minimum cell radius ratio, defined as the ratio of inscribed to circumscribed radii times two~\cite{Caendish1985}, was 0.09.
The mesh within the topmost portion of the part can be seen in Figure~\ref{fig:over}.
The time step utilized for these FE simulations was 0.33~ms.

\section{Pyrometer data versus simulations\label{sec:pvs}}

\subsection{Pyrometer data}

\begin{figure*}[!b]
  \includegraphics[trim=0.0in 0.0in 0.0in 0.0in, clip,width=1\textwidth]{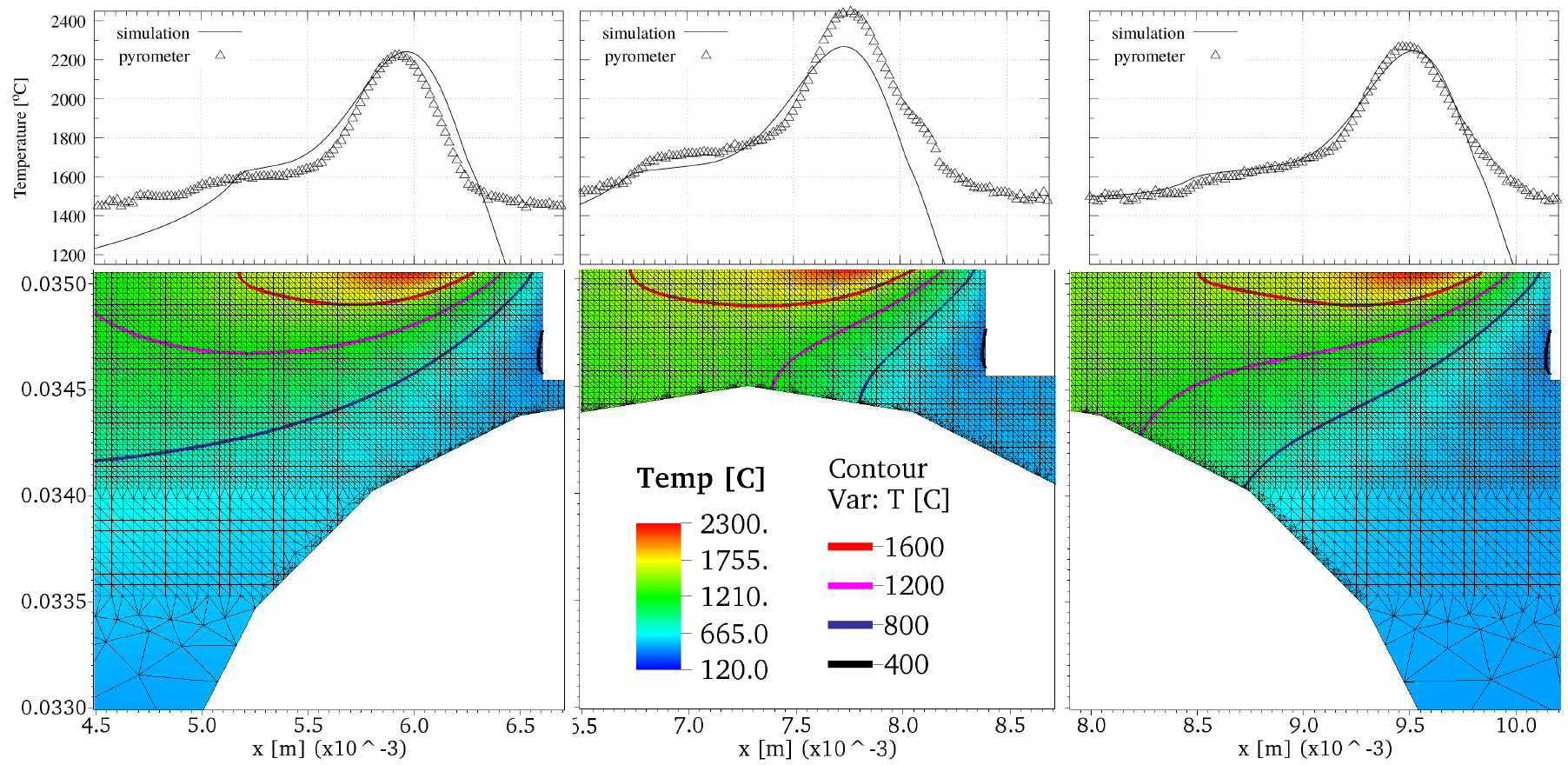}
  \caption{Comparison of simulation results with pyrometer data when depositing above largest void.
    Bottom-center subfigure is a magnified portion of Figure~\ref{fig:upper}.}
  \label {fig:over}
\end{figure*}

\begin{sidewaysfigure*}[!htbp]
  \includegraphics[trim=0.0in 0.0in 0.0in 0.0in, clip,width=1.0\textheight]{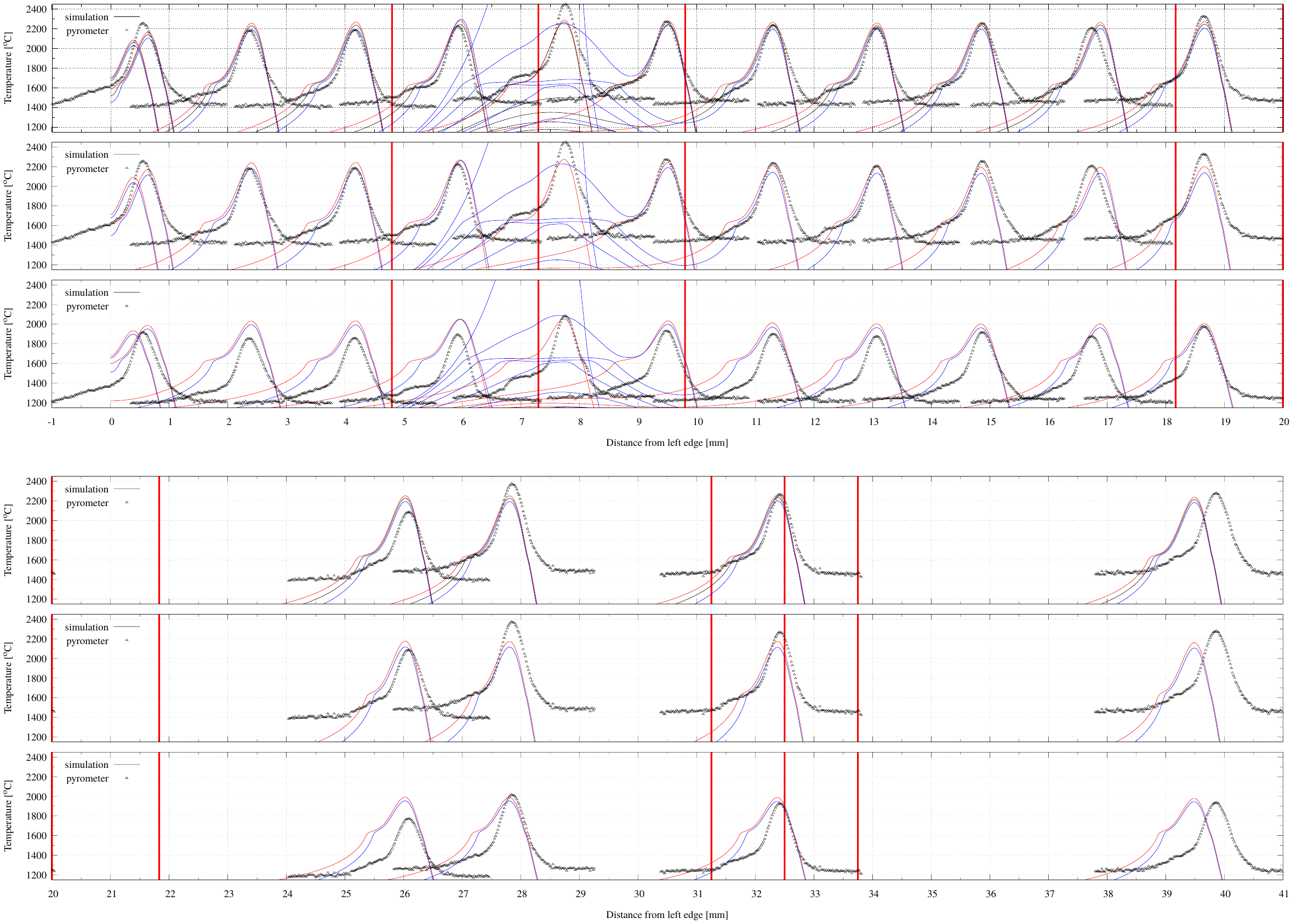}
  \caption{Comparison of the melt pool temperature from simulations
    with pyrometer data. Vertical red lines mark the centers and
    extent of voids. 
    Solid lines show simulated results: the blue line
    shows temperature during deposition of the first layer, the black
    line during deposition of the second layer, and the red line
    during the deposition of the third layer. Triangles show pyrometer measurements.
    Last three plots are
    continuations of the first three plots.}
  \label {fig:over12dir}
\end{sidewaysfigure*}

The present numerical model was calibrated and validated by examining the effects of large geometrical voids on the melt pool size and temperature distribution during the deposition of a part (Figure \ref{fig:geom}).
Pyrometer measurements (Figure~\ref{fig:pyro_maps}) of the melt pool temperature were taken during deposition of the center line of layer 69, which is located just above the largest spherical void.
Each pyrometer scan produced an array of 752 $\times$ 480
temperature values with a grid distance of 12.8~$\mu$m.
Melt pool traveled diagonally at 225{\textdegree}
  angle w.r.t.\ $+x$-direction in Figure~\ref{fig:pyro_maps}.
Average location of the melt pool center was first obtained as the average
point of highest temperature among the scans in
Figure~\ref{fig:pyro_maps}.
Temperature profiles along 190 points of 45{\textdegree} line section of each scan
passing through the melt pool center were then compared with simulations in Figure \ref{fig:over12dir}.
150 leading edge and 140 trailing edge points of each temperature
profile were cut out as temperatures further from melt pool center
were out of pyrometer temperature calibration range.
Similarly, only central portion of pyrometer scans is shown in Figure~\ref{fig:pyro_maps}.
Spatial placement of the pyrometer temperature profiles in
Figure~\ref{fig:over12dir} was set according to pyrometer timestamps
and deposition velocity. Profiles are spaced irregularly
because of variability in the pyrometer sampling rate. The first and last
snapshots of the layer extend slightly out of build surface due to
laser acceleration and deceleration.

\subsection{Simulations\label{sec:simulations}}

Accurate simulations of the direct laser deposition process face several challenges:
\begin{itemize}[wide,labelwidth=!,labelindent=0pt]
\item
A wide range of phenomena are involved: heat conduction, convection, absorption of electromagnetic energy, and phase change. Each of these phenomena needs to be approximated by a numerical model and parameterized, and parameters are temperature dependent \cite{Boivineau2006}.
\item
The meshing process, required by FE method, is non-trivial for an arbitrary complex domain \cite{Dantin2018}. The presence of a large temperature gradient demands a fine mesh, especially with narrow beam radius. A high cooling rate requires a short time step. Therefore, meshing and calculations, especially for 3D domains, are computationally demanding.
\end{itemize}

Material parameters in the present simulations
(Table~\ref{tab:params}) were assumed to be temperature
independent, aiming to reproduce temperature profile above voids with minimal set of parameters.
Two parameters were calibrated to match melt pool
length and maximum melt pool temperature to pyrometer measurements: the effective laser power absorption
coefficient $\alpha$ and the initial temperature of the build $T_i$.
The melt pool length was deduced from the length of the plateau region
extending horizontally from the solidus temperature at the trailing
edge of the temperature profile
\cite{Yadroitsev2014,Cheng2014,Peralta2016,Cheng2017,heigel2018measurement}.

Four laser passes were performed in each simulation. Each simulation
began with a single bare-laser heating pass followed by a deposition
of three layers. The heating pass aimed to achieve temperature
distribution similar to the fully continuous build process which was
used to obtain pyrometer measurements.
The first deposited layer included the top-most portion of the largest
void. The bottom boundary of the second deposited layer is located just
above the top surface of the largest void. The third layer is
deposited on the top of the second layer.

Three cases were simulated to examine the effects of direction of
deposition and laser power. The calculated melt pool surface temperatures from
these three simulations are presented along with pyrometer measurements 
in Figure~\ref{fig:over12dir}. Solid lines show simulated
results: the blue line shows temperature during 
deposition of the first layer, the black line during deposition of the
second layer, and the red line
during the deposition of the third layer.

In the first case, with results shown in the first and fourth plots in
Figure~\ref{fig:over12dir}, the initial heating pass and successive
deposition of three layers were all performed from left to right.
Initial temperature of the build surface
$T_i$ was set to  127~\textdegree C after a few trial-and-error
adjustments in 50~\textdegree C steps to fit the measured melt pool width,
while the laser power absorption coefficient $\alpha$ was set to
$7.2\times10^{-3}$ to fit the maximum melt pool temperature.
Mesh configuration and
calculated temperature of the upper portion of the part during 
deposition of the second layer are shown in Figure~\ref{fig:over}.
As seen in Figure~\ref{fig:over}, the simulated melt pool depth
is smaller than the thickness of the deposited layer. In simulations,
new material is deposited in narrow columns forming a continuous
layer that does not fully melt. This observation
is consistent with powder particles being partially sintered
instead of fully melted \cite{Gong2014}.

In the second case, shown in the second and fifth plots in
Figure~\ref{fig:over12dir}, the simulation began with a single heating
pass from right to left, followed by a deposition of three
layers in alternating directions. That is, the first layer was
deposited left to right, the second right to left, and the third
left to right. The temperature of the build surface
was initialized at $T_i = 30$~\textdegree C. A pause of 240~ms
between depositions of successive layers was introduced in
simulations to compensate for the successive layers being deposited in alternate
directions. The pause is needed to avoid excessive temperature when
starting to deposit a new layer on the top of recently deposited material.
Simulation results from the deposition of the third layer in the
second case (red line in the second plot in
Figure~\ref{fig:over12dir}) were found to be the best match to the
measured data because measurements were taken when successive tracks
were deposited in alternating directions (Fig.~\ref{fig:depo}).

For the first two cases, the pyrometer data in Figure~\ref{fig:over12dir} was scaled up by a multiplicative
factor of 1.18. This adjustment displaced the solidification
plateau from pyrometer data closer to the plateau from simulations,
which is just above the solidus temperature. The need for this
adjustment can be attributed to the present numerical model not accounting for the flow
of inert gas which cools down the build surface,
or inaccuracy of the pyrometer.
Moreover, the 2D model does not allow lateral heat dissipation
perpendicular to the direction of deposition, what results in
overestimation of melt pool temperature.

In the third case, shown in the third and sixth plots in
Figure~\ref{fig:over12dir}, pyrometer data were not scaled up. The
direction of deposition alternated with pauses as in the second case,
however the laser power absorption coefficient was cut by 25\% to match
1.18$\times$ lower maximum melt pool temperature. The initial
temperature of the build surface was increased to $T_i =
427$~\textdegree C to match the melt pool length.

The blue solid lines in Figure~\ref{fig:over12dir} demonstrate very
high temperature during the deposition in the region including top portion of the large
void. This is mainly due to a small amount of cold material being
added with heat supply at the same level as when the large
amount of material is added. In experimental setup, the deposition in
this region would experience distortions with largely unknown
consequences for temperature measurements. The deposition of the 
layer including top portion of the large void is therefore not
considered for comparison with
measured temperature. Instead, simulated results from the deposition of two lines
above the large void are compared with the temperature measured during the
deposition of the center line above the large void.

The experiment was done on a 1~cm thick wall, and the top
of each cavity only spanned one to three line passes in general. This
would be about 0.5~mm to 1.5~mm. Since there would be at least 4~mm on
each side of the cavity that was fully supported, and the cavities
were spherical, we made the assumption that there was enough
supporting material and structure to reduce most of the distortion
that would be seen on the layer just above the cavity. Thus we did not
assume distortion would have an effect on the melt pool temperature
when depositing these lines. However, distortion and residual stress
effects are something we are still looking into with current and
upcoming studies.


%

\section{Conclusions}

Direct laser deposition of a narrow overhang layer, as the thin layer above a large round void presented in this work, is a challenge for additive manufacturing due to insufficient support of newly deposited material.
The two-dimensional model of the temperature evolution during direct laser deposition developed in this work was applied to examine the melt pool size and temperature during the build process above large geometrical voids.
Trailing edge of the longitudinal section of the temperature profile across the melt pool exhibited a plateau region extending horizontally from near-solidus temperature. The numerical model was calibrated by matching the length of solidification plateau between the simulation and pyrometer data. Implications of the study are:
\begin{itemize}[wide,labelwidth=!,labelindent=0pt]
    \item A 2D model can provide a reasonable approximation of the central longitudinal section of the melt pool temperature profile.
    \item The solidification plateau in the melt pool temperature profile increases in length when depositing material above large void.
    \item The portion of the part above large voids cools down slower. It remains at the temperature higher than the regions without underlying voids. The effect is more pronounced with larger voids.
    \item The maximum temperature of the melt pool increases during deposition above a large void. A smaller increase of maximum melt pool temperature was observed above small voids.
    \item Deposition in alternate directions leads to increased maximum melt pool temperatures at the time when the direction of deposition is reversed. A pause between the deposition of successive layers helps to mitigate this effect. Deposition in a single direction leads to more uniform maximum melt pool temperatures.

    \item Melt pool length, as characterized by extent
      of the solidification plateau in the melt pool temperature
      profile, correlates strongly with temperature of build
      surface, while the maximum melt pool
      temperature correlates with power of the heat source.
\end{itemize}

The present 2D FE model of the DED process can, upon calibration,
closely reproduce the melt pool shape during the deposition of
geometrically complex parts. The two calibrated parameters are the effective heat
absorption coefficient for laser power and initial temperature of the
build surface. With serial computational
time in the order of tens of hours, this model can estimate effects of
process parameters on the melt pool temperature profile.
Aiming at minimal complexity needed to reproduce the measured melt
pool temperature variation, the model is amendable to improvements.
Changes as making material properties temperature-dependent and
weighting the thermal properties by the phase fractions instead of
using average properties are expected to improve physical
representation and are in plans for the future work.
Relying on
the aid of parallel implementation, which is already partially
supported by FEniCS framework, the model can be extended to
realistic-scale 3D DED process simulations.

\section*{Acknowledgment}

Research was sponsored by the Army Research Laboratory and was
accomplished under Cooperative Agreement Number W911NF-15-2-0025. The
views and conclusions contained in this document are those of the
authors and should not be interpreted as representing the official
policies, either expressed or implied, of the Army Research Laboratory
or the U.S. Government. The U.S. Government is authorized to reproduce
and distribute reprints for Government purposes notwithstanding any
copyright notation herein.
%


\bibliographystyle{model1-num-names}
\bibliography{bb}







\appendix

\section{Pyrometer maps}

\begin{figure*}[!b]
  \center
  \includegraphics[trim=0.0in 0.0in 0.0in 0.0in, clip,height=.97\textheight]{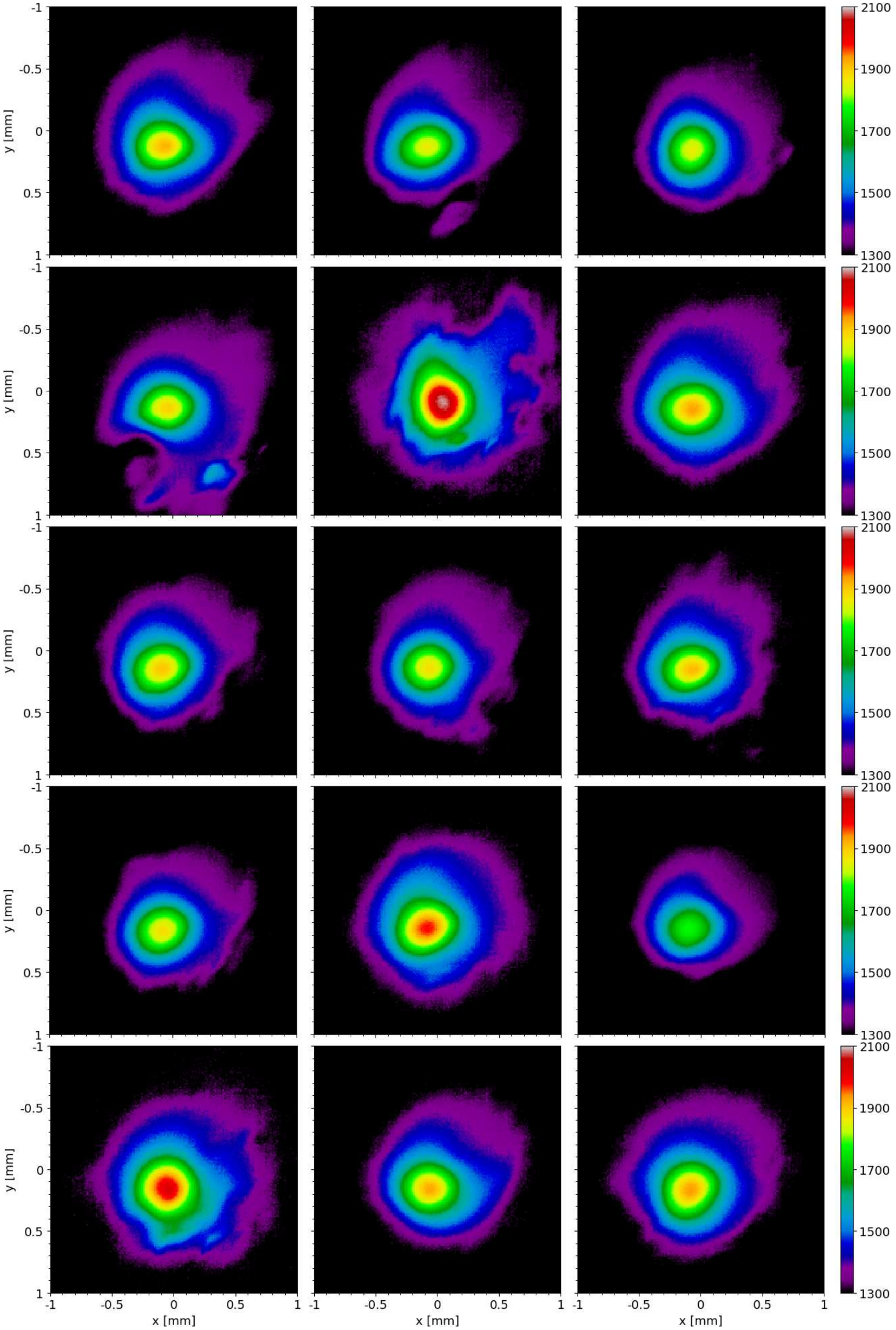}
  \caption{Pyrometer scans of the melt pool temperature taken from a layer just above the largest circular void. Snapshot time stamp increases along the rows. Melt pool travels in the south-west diagonal direction.}
  \label{fig:pyro_maps}
\end{figure*}

\end{document}